# Stress analysis of functionally graded hyperelastic variable thickness rotating annular thin disk: A semi-analytic approach

Ehsan Jebellat[1*], Iman Jebellat[2]

1- School of Mechanical Engineering, College of Engineering, University of Tehran, Tehran, Iran
2- Department of Mechanical Engineering, Sharif University of Technology, Tehran, Iran

Corresponding author: Ehsan.Jebellat@ut.ac.ir

**Abstract**

Functionally graded materials (FGMs) represent a promising class of advanced materials designed with tailored microstructures to achieve optimized mechanical, thermal, and functional properties across varying gradients. The strategic integration of distinct materials within functionally graded materials offers engineers unprecedented control over properties such as strength, thermal conductivity, and corrosion resistance, enabling innovative solutions for demanding applications in aerospace, automotive, and biomedical industries. This study investigates a rotating annular thin disk with variable thickness composed of incompressible hyperelastic material, made up of functionally graded properties under large deformations. To elucidate these phenomena, a power relation is employed to delineate the changes in cross-sectional geometry $m$, the material property $n$, and the angular velocity $\omega$ of hyperelastic material. Constants used for hyperelastic material are obtained from the experimental data. Equations are solved semi-analytically for different values of $m$, $n$, and $\omega$, and the values of radial stresses, tangential stresses, and elongation are calculated and compared for different conditions. Results show that thickness and FG properties have a significant impact on the behavior of disk, so that the expected behavior of the disk can be obtained by an optimal selection of the disk's geometry and material properties. By selecting the optimum values for these variables, the location of maximum stress can be controlled in large deformations, thereby furnishing significance advantages in structural design and material selection.

## 1. Introduction

Hyperelastic materials can endure high values of strain and, in fact, can be exposed to large deformations and remain fully elastic without losing their main properties. In general, the behavior of hyperelastic materials changes nonlinearly and their structures are based on hyperelastic models [1]. Different classes of materials such as elastomers, polymers, and foams have the potential of large hyperelastic deformations. Rubbers are widely used as vibration isolator, energy-saving parts in the automotive industry and shield and buffer in the parts exposed to impact loads [2]. In hyperlastic materials, the ratio of the secondary length to the primary length is usually between 5 to 10 and the stress-strain curve is nonlinear, so the material does not follow Hooke's law [3]. For small tensions, one can define the slope of the curve as the modulus of elasticity, which is about 1 MPa. High extensibility and low modulus of elasticity of rubbers, compared to solids, such as metals, which have the high value of modulus of elasticity (about 200 GPa) and a low extensibility (1.01), makes a remarkable difference between rubbers and solids like metals.

To study the structural equations of the rubbers, two types of phenomenological viewpoints are used; a viewpoint based on the continuum mechanics, and a viewpoint of the statistical and kinetics methods. So far, many structural equations have been derived for the modeling of the hyperelastic behavior of rubber materials based on the continuum mechanics. For this purpose, Ogden [4] presented an energy function based on the strain energy density functions for nonlinear behavior of rubber materials with large deformations, in which there was a series of coefficients and constants that the specific state relations of this model, such as Mooney-Rivlin, Neo-Hookean, and Varga models were introduced. Mooney [5] proposed the large deformations theory, and Rivlin [6] developed this theory for rubbers. Yeoh [7] introduced an energy function to describe the behavior of volcanized carbon-black-filled rubbers. Arruda and Boyce [8] presented a statistical model whose parameters were physically related to the chain orientation, including the deformation of the structure of the three-dimensional rubbers grid. Blatz and Ko [9] presented a strain energy function for compressive foamed-elastomers, which was a combination of theoretical argumentations and laboratory data. Beatty [10], Horgan and Polignone [11], and Attard [12] have also introduced new structural models for rubbers.

Material design in one of the state-of-the-art topics since it helps making various types of materials and structures for different applications. In this case, smart materials and structures are one of the most novel topics in the literature of materials science [13]. Several researches have been done on shape-memory materials [14, 15] and smart composites [16]. Most materials existed in nature are not homogeneous and are considered homogeneous only for simplification. Heterogeneous materials have different types. The use of FGMs is one of the practical hypotheses to consider the effect of heterogeneity [17]. FGMs exhibit different properties in different regions due to the gradual change of chemical composition, distribution and orientation or the size of the reinforcing phase in one or more dimensions [18]. FGMs have been used in many fields such as aerospace [19], energy [20], biology [21], electro-magnetism [22], controllers [23], network [24, 25], vehicles [26, 27], and other fields [28, 29] through the ingenious combination of inorganic and organic materials such as metals, ceramics, and plastics. This gradual change in structure and properties has led to the expansion of the application of these materials, especially in cases requiring different properties in different regions. FGMs were initially introduced in 1984 by a group of scientists at the University of Sendai, Japan [30], since then extensive researches were carried out on these materials. Due to the continuity of their mechanical, thermal, and magnetic properties in FGMs, stresses and gradients are continuous, increasing their strength. These gradual changes in properties of the structure of FGMs cause the consistent strength between different layers of them. However, in laminated composite materials, the interference between fiber structures causes inconsistency in mechanical properties. In the realm of materials science, stress analysis has been conducted on different materials to investigate their fracture properties in nanoscale [31] or failure response in macroscale [32]. Determining material parameters to describe mechanical behavior stands as another outcome of stress analysis [33]. Moreover, the effect of geometric parameters on the structural properties which governs the mechanical behavior of a system is investigated using stress analysis [34].

Two worldwide Symposium on heterogeneous FGMs was held in Sendai in 1990 and in San Francisco in 1992, leading to extensive research in this field [35]. The studies of structures made from heterogeneous FGMs over the past decade have been of interest to many researchers. For example, Sankar [36] provided the elasticity solution for an Euler-Bernoulli beam under static transverse loadings. In another study, Benatta [37] presented an analytical solution for heterogeneous FGM beams under a bending loading. To understand and predict the nonlinear behavior of natural and artificial materials, such as rubbers, foams, and tissues of living organisms with nonlinear elastic behavior and some degrees of heterogeneity, it is necessary to analyze their structural equations to be utilized in simulation. Heterogeneous hyperelastic FGMs have hyperelastic behavior, and their mechanical properties change continuously from one point to another in the specified direction. In other words, these materials gradually transform from one to another material. First, FGM rubbers were manufactured by Ikeda [38] in the laboratory. Bilgili et al. [39] investigated the heterogeneity effects on rubber parts under tensile and shear loadings. In another study, Bilgili et al. [40] investigated the effects of thermal loading on heterogeneous rubber parts. Batra [41] also has investigated the behavior of nonlinear heterogeneous thick-walled cylinders with large deformations using numerical methods. Reddy [42] made an analysis of elastic shells. Shells are the main structural elements in many engineering applications, including pressure vessels, submarines, ships, aircraft wings and bodies, car tire tubes, rockets, concrete roofs, chimneys, cooling towers, and liquid storage tanks. In addition, they can be found in nature in the form of leaves, eggs, inner ears, bladders, blood vessels, skulls, and geological structures.

FGM rubber tubes in the form of a finite thermoelastic model with a molecular pattern, was investigated by Bilgili [43]. The heterogeneity role in the deformation of elastic bodies was investigated by Saravanan and Rajagopal [44]. In another study, a comparison between the response of heterogeneous isotropic spherical and cylindrical shells and their homogeneous counterparts was performed by Saravanan and Rajagopal [45]. One year later, the same authors analyzed the expansion and torsion of a heterogeneous incompressible elastic circular cylinder [46]. Iaccarino and Batra analyzed the expansion and radial contraction of a hollow sphere made of second-order heterogeneous compressible isotropic elastic materials with two material parameters [47]. Batra analyzed the torsion of a cylinder made of incompressible Hookean materials with a variable shear modulus along the axial direction and calculated the axial variations of shear modulus for controlling the torsion angle of cross section [48]. Batra and Bahrami investigated a pressure vessel made of FGM rubber, such as pressurized materials [49]. In order to find the stress components of the pressure vessel, they assumed an axially symmetrical radial deformation of a circular cylinder, which was composed of FGM Mooney-Rivlin materials with the parameters changing continuously and radially by power law or relativity. They found that in order to make the power law component equal to 1, tangential stress for the cylinder under the internal pressure should be uniform in the whole cylinder. In another study, Batra presented general relations for axial symmetric deformations in FGM cylinders and rubber spheres [50]. It is worth mentioning that several analysis based on machine learning techniques [51, 52], reinforcement learning strategy [53, 54] and dynamic programing methods [55, 56] can be performed to predict the fatigue life of FGM specimen [57], find static structural reliability of FGM frames [58], calculate cutting forces in milling processes of FGM [59], and so many other applications.

By reviewing the related works, it was found that no research has been performed in the field of analysis of hyperelastic FGM rotary disk with variable cross section. Therefore, in this paper, we analyze a rotary disk consisting of FGM materials with variable cross section and also hyperelastic behavior. The energy function used in this paper is the Neo-Hookean function. The cross section of disk changes based on power law with the disk radius. The results in this paper are presented for different values of FGM material component, cross-section component, and angular velocity of the disk. The outline of this paper is as follows: The first section introduced the related context of FGM and reviewed the literature. Section 2 presents the analytical solutions, and section 3 investigates the results and discussion. The effects of material properties, cross-sectional area, and angular velocity are discussed in this part. Finally, important remarks and conclusions are mentioned in section 4 of this article.

## 2. Formulation

Figure 1 shows an FGM hyperelastic annulus disk with variable thickness. The inner radius and outer radius of the disk are shown as *A* and *B*, respectively. The hyperelastic material is considered to be fully incompressible. The non-deformed and deformed shapes of the disc are shown in the reference coordinate system $(R, \Theta, Z)$ and current coordinate system $(r, \theta, z)$, respectively. The length of the hypotenuse is *r*, and *θ* is the measure of the angle formed by the positive x-axis and the hypotenuse. The z-coordinate describes the location of the point above or below the xy-plane.

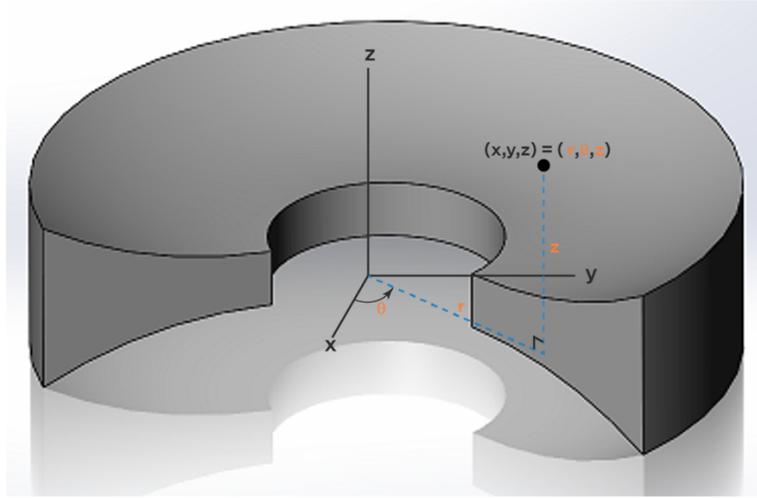

**Figure 1.** Cross section of a disk with variable thickness.

Because of symmetry, the deformation function of the rotating disk can be expressed as,

$$\begin{cases} r = f(R,Z) \\ z = g(R,Z) \\ \theta = \Theta \end{cases} \qquad (1)$$

Therefore, the deformation gradient tensor can be expressed as follows.

$$F = \begin{bmatrix} \dfrac{\partial f}{\partial R} & 0 & \dfrac{\partial f}{\partial Z} \\ 0 & \dfrac{r}{R} & 0 \\ \dfrac{\partial g}{\partial R} & 0 & \dfrac{\partial g}{\partial Z} \end{bmatrix} \qquad (2)$$

For an incompressible hyperelastic material, Cauchy's stress can be shown as [60]

$$\boldsymbol{\sigma} = -p\mathbf{I} + 2W_1\mathbf{B} + 2W_2(I_1\mathbf{B} - \mathbf{B}^2) \qquad (3)$$

So that in the above equation, $p$ is the constraint pressure existed on the problem and indicates the incompressibility of material and is obtained from the boundary conditions of the problem. Also, $B$ is the left Cauchy-Green tensor shown as follows.

$$B = FF^T = \begin{bmatrix} \left(\dfrac{\partial f}{\partial R}\right)^2 + \left(\dfrac{\partial f}{\partial Z}\right)^2 & 0 & \left(\dfrac{\partial f}{\partial R}\right)\left(\dfrac{\partial g}{\partial R}\right) + \left(\dfrac{\partial f}{\partial Z}\right)\left(\dfrac{\partial g}{\partial Z}\right) \\ 0 & \left(\dfrac{r}{R}\right)^2 & 0 \\ \left(\dfrac{\partial f}{\partial R}\right)\left(\dfrac{\partial g}{\partial R}\right) + \left(\dfrac{\partial f}{\partial Z}\right)\left(\dfrac{\partial g}{\partial Z}\right) & 0 & \left(\dfrac{\partial g}{\partial R}\right)^2 + \left(\dfrac{\partial g}{\partial Z}\right)^2 \end{bmatrix} \quad (4)$$

$I_1$, $I_2$, and $I_3$ are the first, second, and third invariants of the left Cauchy-Green tensor, respectively.

$$I_1 = tr(\mathbf{B}) = \left(\dfrac{\partial f}{\partial R}\right)^2 + \left(\dfrac{\partial f}{\partial Z}\right)^2 + \left(\dfrac{r}{R}\right)^2 + \left(\dfrac{\partial g}{\partial R}\right)^2 + \left(\dfrac{\partial g}{\partial Z}\right)^2 \quad (5)$$

$$I_2 = \dfrac{1}{2}(tr(\mathbf{B}^2) - tr(\mathbf{B})^2) \quad (6)$$

$$I_3 = \det(\mathbf{B}) \quad (7)$$

Also, $W_1$ and $W_2$ represent the energy derivatives relative to the first invariant ($I_1$) and the second invariant ($I_2$), respectively.

$$W_1 = \dfrac{\partial \psi}{\partial I_1} \quad (8)$$

$$W_2 = \dfrac{\partial \psi}{\partial I_2} \quad (9)$$

Now, with the expansion of the Cauchy's stress relation of equation (3) and by abandoning the shear stress, following relation is obtained:

$$\dfrac{\partial f}{\partial Z} = 0 \quad (10)$$

$$\dfrac{\partial g}{\partial R} = 0 \quad (11)$$

Therefore, the deformation gradient tensor can be written as follows:

$$F = \begin{bmatrix} \dfrac{\partial f}{\partial R} & 0 & 0 \\ 0 & \dfrac{r}{R} & 0 \\ 0 & 0 & 1 \end{bmatrix} \quad (12)$$

Given the assumption of the incompressibility of material, the following relation can be written:

$$J = \det(\mathbf{F}) = \dfrac{dr}{dR}\dfrac{r}{R} = 1 \quad (12)$$

Considering the internal radius of the disk (A) and the external radius of the disk (B), their deformed shapes can be written as follows:

$$\begin{cases} a = f(A) \\ b = f(B) \end{cases} \tag{13}$$

Therefore, considering the boundary conditions above, equation (12) is solved as follows:

$$r^2 = R^2 - B^2 + b^2 \tag{14}$$

By defining $\lambda_{\theta(B)} = \dfrac{b}{B}$ as the elongation of the outer radius and $\beta = \dfrac{B}{A}$, the elongation at any point can be expressed as follows:

$$\lambda_\theta = (1 - \dfrac{1}{R^2}(B^2 + b^2))^{\dfrac{1}{2}} \tag{15}$$

The Cauchy's stress relation can be written as follows:

$$\begin{bmatrix} \sigma_r & 0 & 0 \\ 0 & \sigma_\theta & 0 \\ 0 & 0 & \sigma_z \end{bmatrix} = -p \begin{bmatrix} 1 & 0 & 0 \\ 0 & 1 & 0 \\ 0 & 0 & 1 \end{bmatrix} + 2W_1 \begin{bmatrix} \left(\dfrac{dr}{\partial R}\right)^2 & 0 & 0 \\ 0 & \left(\dfrac{r}{R}\right)^2 & 0 \\ 0 & 0 & 1 \end{bmatrix}$$

$$+ 2W_2 \left( \left[\left(\dfrac{dr}{\partial R}\right)^2 + \left(\dfrac{r}{R}\right)^2 + 1\right] \begin{bmatrix} \left(\dfrac{dr}{\partial R}\right)^2 & 0 & 0 \\ 0 & \left(\dfrac{r}{R}\right)^2 & 0 \\ 0 & 0 & 1 \end{bmatrix} - \begin{bmatrix} \left(\dfrac{dr}{\partial R}\right)^4 & 0 & 0 \\ 0 & \left(\dfrac{r}{R}\right)^4 & 0 \\ 0 & 0 & 1 \end{bmatrix} \right) \tag{16}$$

The energy function considered in this paper is Neo-Hookean. Since the Neo-Hookean constant coefficient is equal to the shear modulus, it can be considered as FGM with considering this coefficient to be variable relative to the initial radius of the object.

$$W = \dfrac{\mu}{2}(I_1 - 3)^\alpha \tag{17}$$

Also, FGM changes has been considered as follows and the changes of the FGM properties relative to the function power has been plotted in the figure 2.

$$\phi = \phi_0 (\frac{R}{A})^n \tag{18}$$

Therefore, changes of the shear modulus relative to the radius has been expressed as follows:

$$\mu = \mu_0 (\frac{R}{A})^n \tag{19}$$

in which $R$ represents the initial radius of the disk, $A$ is equal to the radius of the disk, and $n$ represents the changes in module of the disk. By merging the equations (14) and (13), stresses are obtained as follows:

$$\sigma_r = -p + \alpha \mu_0 (\frac{R}{A})^n (I_1 - 3)^{\alpha-1} \left(\frac{dr}{\partial R}\right)^2 \tag{20}$$

$$\sigma_\theta = -p + \alpha \mu_0 (\frac{R}{A})^n (I_1 - 3)^{\alpha-1} \left(\frac{r}{R}\right)^2 \tag{21}$$

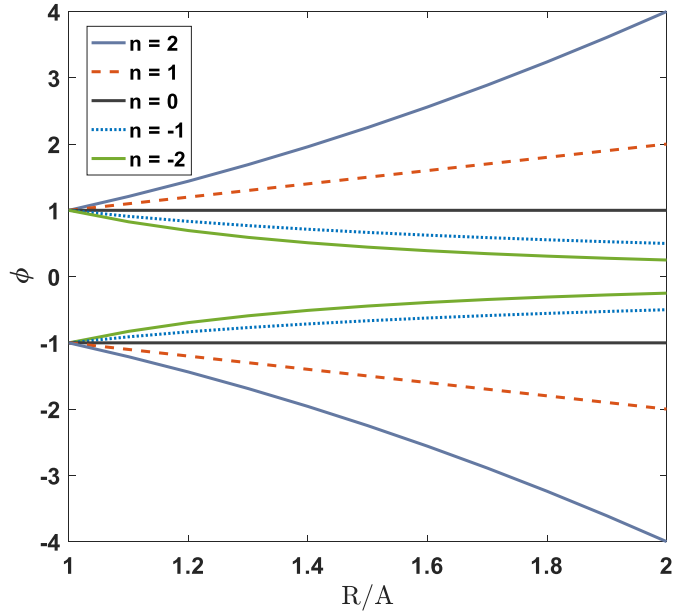

**Figure 2.** Changes in FGM properties for different values of material property $n$.

The governing equilibrium equation of the disk with variable thickness is as follows, in which $z$ indicates the function of the thickness variation.

$$\frac{d\sigma_r}{dr} + \frac{1}{z}\frac{dz}{dr}\sigma_r + \frac{\sigma_r - \sigma_\theta}{r} = -\rho r \omega^2 \tag{22}$$

Thickness variations are shown as the function below, and figure 3 represents the thickness variations relative to different values of the power component of the function.

$$Z = C_0 \left(\frac{R}{A}\right)^m \tag{23}$$

Now, considering the function of thickness variations and density variations, equation (22) is rewritten as follows:

$$\frac{d\sigma_r}{dr} + \frac{mr}{R^2}\sigma_r + \frac{\sigma_r - \sigma_\theta}{r} = -\rho_0 \left(\frac{R}{A}\right)^n r\omega^2 \tag{24}$$

Inserting equations (20-21) into equation (24), it can be rewritten as:

$$\frac{d\sigma_r}{dr} + \frac{mr}{R^2}\sigma_r + 2\frac{W_1}{r}\left[\left(\frac{dr}{\partial R}\right)^2 - \left(\frac{r}{R}\right)^2\right] = -\rho_0 \left(\frac{R}{A}\right)^n r\omega^2 \tag{25}$$

By inserting the equation (14) into the Equation (25) we have:

$$\frac{d\sigma_r}{dr} + \frac{mr}{r^2 - b^2 + B^2}\sigma_r + \frac{\alpha\mu_0}{r} \frac{R^n}{(r^2 - b^2 + B^2)^{\frac{n}{2}}} \left(\frac{r^2}{r^2 - b^2 + B^2} + \frac{r^2 - b^2 + B^2}{r^2} - 2\right)^{\alpha-1}$$

$$\times \left(\frac{r^2 - b^2 + B^2}{r^2} - \frac{r^2}{r^2 - b^2 + B^2}\right) = -\rho_0 \left(\frac{R}{A}\right)^n r\omega^2 \tag{26}$$

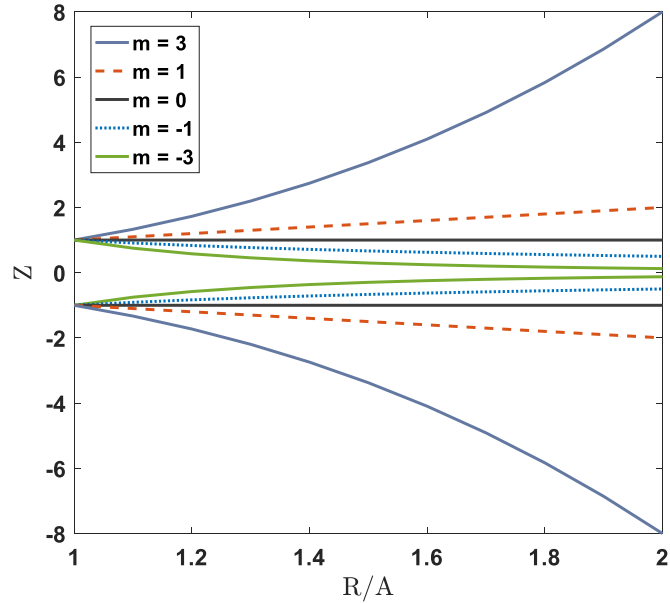

**Figure 3.** Thickness variations of the disk for different values of geometry *m*.

To solve the above differential equation numerically, the fourth order Runge-Kutta method and the following two boundary conditions - which represent the values of stresses in the inner and outer radius of the disk after deformation - have been used.

$$\begin{cases} \sigma_r(r=a)=0 \\ \sigma_r(r=b)=0 \end{cases} \quad (27)$$

## 3. Results and discussion

Constant values of $\beta=3$, $\mu_0$, $\rho_0$ have been obtained using experiments performed by Batra. So that the values are $\alpha=1.5$, $\mu_0=0.0521 MPa$ and $\rho_0=920 Kg/m^3$ [61]. In order to solve the above equations, a disk with the inner radius of $A=1$ and the outer radius of $B=2$ has been exposed to the rotation with angular velocities of $\omega=3$, $\omega=4.5$, and $\omega=6$ rad/s. The considered FGM parameters and geometries are in the forms of $n=2,1,0,-1,-2$ and $m=3,1,0,-1,-3$, respectively.

Since the neo-Hookean energy function possesses linear behavior in small deformations, the results of present study were compared with the results of a linear elastic disk with a variable cross section, provided by Timoshenko and Goodier, in low rotational speeds [62]. The results show a very good agreement between the results of present study and Timoshenko's results.

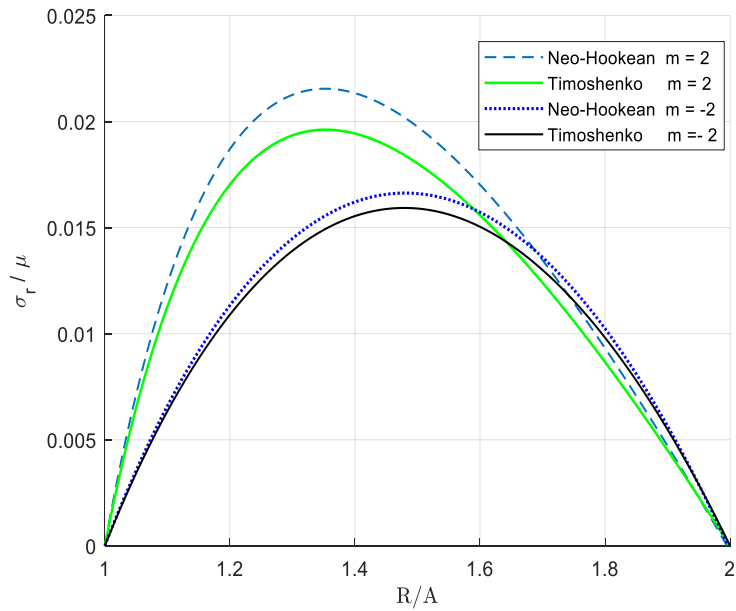

**Figure 4.** Comparison of radial stress distribution of neo-Hookean and Timoshenko models for angular velocity $w=1.5$ rad/s.

### *3.1. Effects of material properties*

In order to solve the above equations, a disk with the inner radius of $A=1$ and the outer radius of $B=2$ has been exposed to a rotation with angular velocities of $\omega=3$, $\omega=4.5$, and $\omega=6$ rad/s. As mentioned, the considered FGM parameters and geometries are in the forms of

$n = 2, 1, 0, -1, -2$ and $m = 3, 1, 0, -1, -3$, respectively. Results are provided for $\beta = 2$. Figure 5a shows the radial stress distribution for angular velocity of $\omega = 3$ rad/s and in a constant cross section for different values of *n*. The radial stress of the disk has been increased by increasing the value of *n*. At a constant cross section, with increasing *n*, the maximum radial stress developed in the disk moves toward the outer radius, which can be considered from the viewpoint of design. Figure 5b shows the tangential stress distribution on the disk. Values of tangential stress in the inner radius of the disk is the maximum value of the disk, so that as we approach the outer radius, the tangential stress reduces, so that in the outer radius of the disk for $n = -2.1$, the value of tangential stress approaches to zero. The reduction of tangential stress in the disk is such that with increasing *n*, the reduction trend is linear, but the reduction trend is nonlinear for negative values of *n*.

Figure 5c indicates the amount of elongation along the disk. As approaching the end of the disc, the elongation has been reduced so that the maximum elongation has been occurred in the inner radius. As the value of *n* increases, the elongation in the disk has been increased. This difference is such that, when approaching the end of the disk, differences in elongations in different values of *n* are less relative to each other, but differences in the values of *n* cause more differences of elongations in the inner radius.

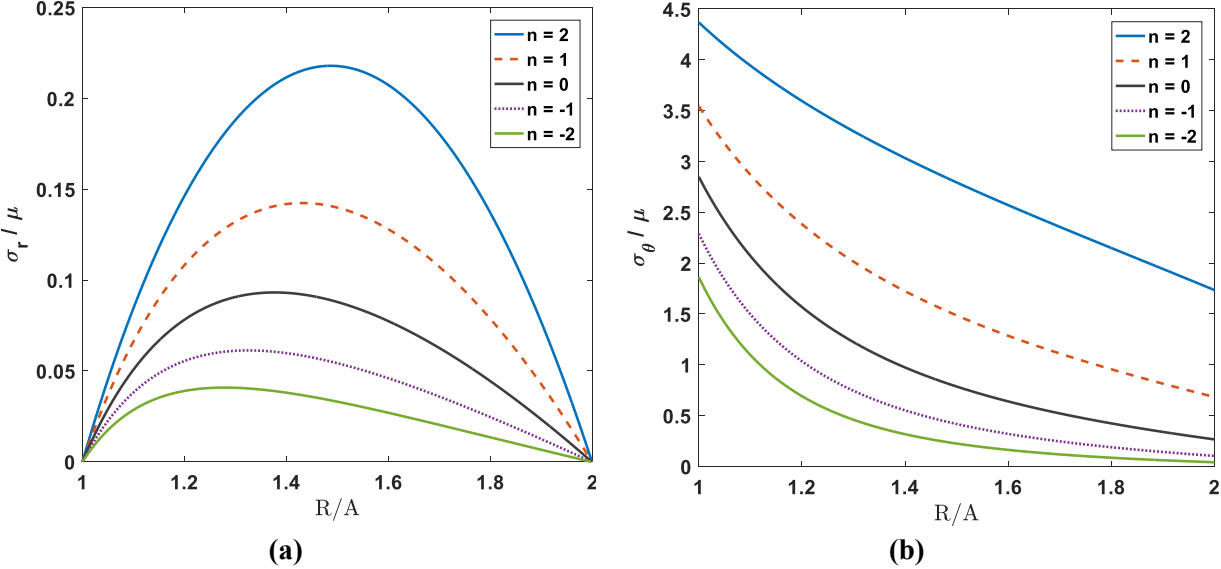

(a)                                                (b)

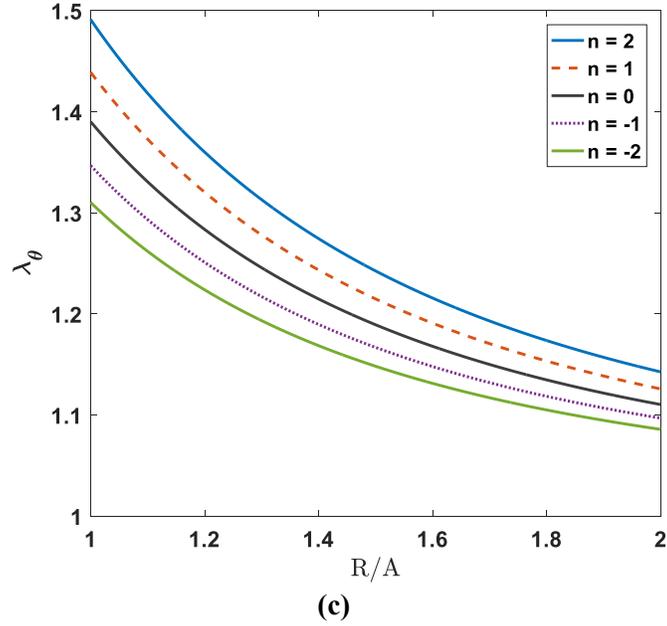

**(c)**

**Figure 5.** Results for $w = 3$ rad/s and $m = 0$ and different values of $n$, **(a)** radial stress distribution, **(b)** tangential stress distribution, **(c)** elongation.

Figure 6a indicates the radial stress distribution for $m = -3$. The values of the obtained stresses are less than their values in $m = 0$ state, that can be due to the reduction of rotating material. The maximum radial stress towards more to the end of the disk than the previous state, so that at $n = 2$, the maximum value has been occurred in $\frac{R}{A} = 1.6$, while in the $m = 0$ state, the maximum value has been occurred in $\frac{R}{A} = 1.5$. Figure 6b shows the tangential stress of the disk, which has less values than its values in $m = 0$ state. Figure 6c indicates the value of elongation along the disk, which is less than its values in m = 0 state, and as in the previous state, the elongation of the disk has been reduced by approaching to the end of the disk so that the highest value of elongation has been occurred in the inner radius.

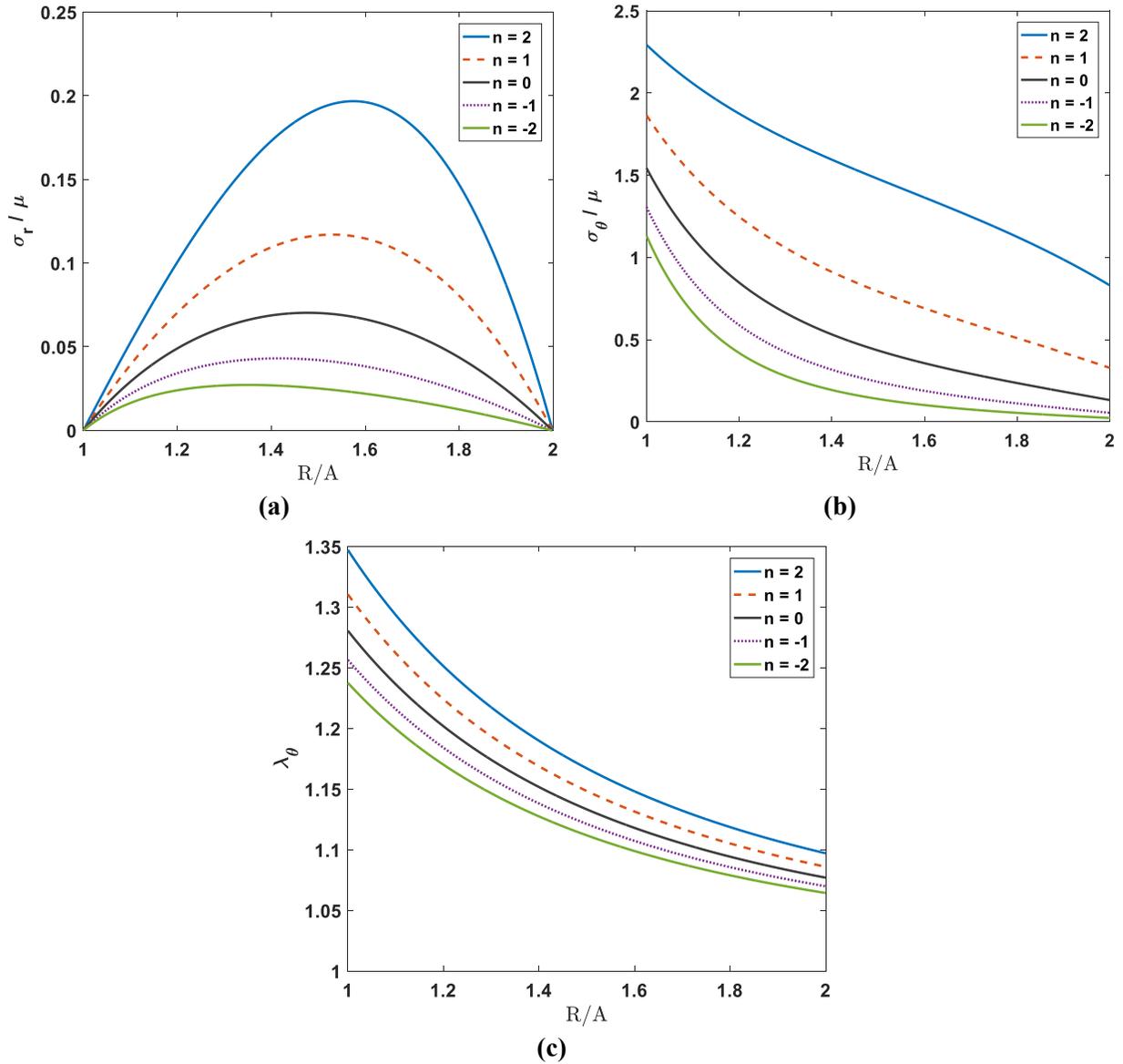

**Figure 6.** Results for *w* = 3 rad/s and *m* = -3 and different values of *n*, **(a)** radial stress distribution, **(b)** tangential stress distribution, **(c)** elongation.

Figure 7a shows the radial stress for *m* = 3 and different values of *n*. Such as the previous figures, the stress is increased with increasing *n*, which can be due to the increasing trend of thickness of the disk. In this figure, the maximum radial stress towards more to the inner radius relative to the two previous states. Figure 7b shows the radial stress of the disc. The value of radial stress on the disk in this case is higher than the previous two states. Figure 7c indicates the elongation, which in this case is higher than the previous two states, and such as the two previous states, by approaching to the end of disk, elongations become more converged.

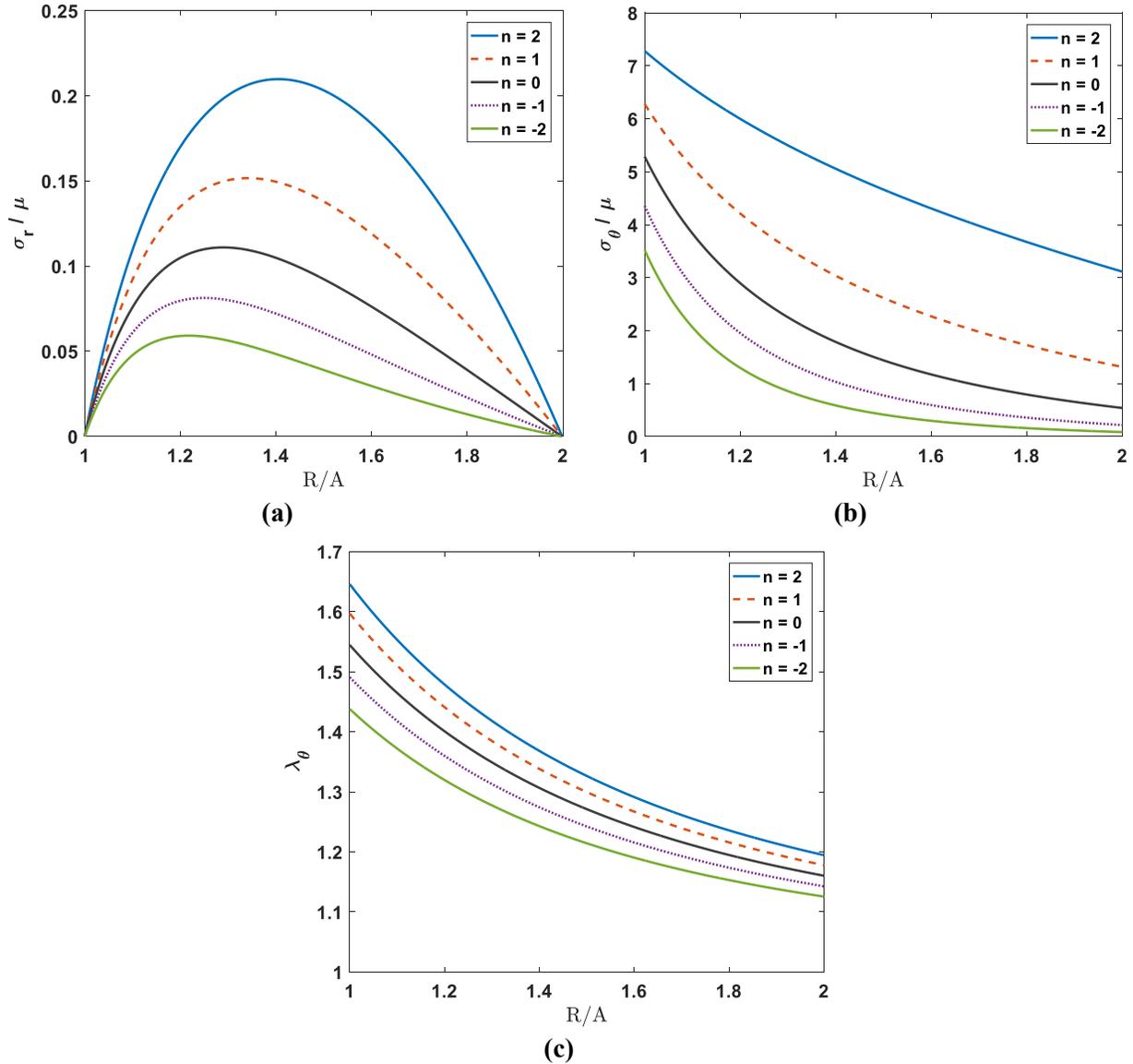

**Figure 7.** Results for *w* = 3 rad/s and *m* = 3 and different values of *n*, **(a)** radial stress distribution, **(b)** tangential stress distribution, **(c)** elongation.

### *3.2. Effects of cross-sectional area*

Figure 8a shows the radial stress distribution for angular velocities of $\omega = 4.5$ rad/s and *n* = 0 for different values of *m*. As *m* increases, the maximum stresses on the disk increase and move toward the inner radius. However, this increase in *m* does not mean more stresses in the entire radius of disk, so that, by approaching to the end of the disk, stresses are in decreasing trend by increasing *m*. In figure 8b, it has been shown that the tangential stress increases by increasing the values of *m* and decreases by approaching to end of the disk. Figure 8c represents the elongation along the disk, which increases with increasing the values of *m*.

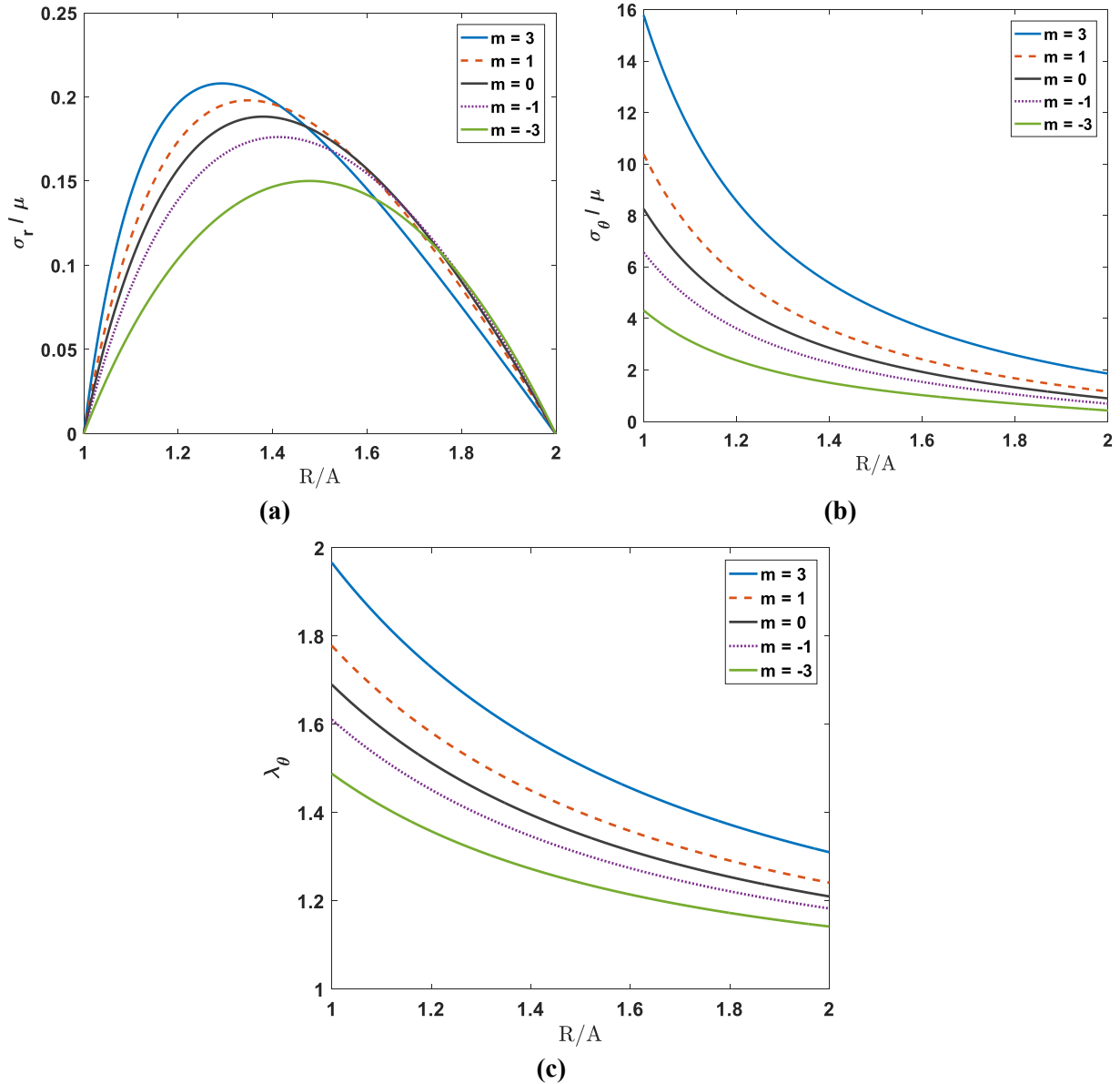

**Figure 8.** Results for $w = 4.5$ rad/s and $n = 0$ and different values of $m$, **(a)** radial stress distribution, **(b)** tangential stress distribution, **(c)** elongation.

Figure 9a shows the radial stress distribution for $n = -2$ in different cross sections, in which the values of stress have been decreased relative to the previous state. By increasing the value of $m$, the maximum stress increases and moves towards the inner radius. Such as the $n = 0$ state, this increase is not always true along the entire radius of disk, so that, by approaching to the end of the disk and in $\frac{R}{A} = 1.8$, this trend changes and the values of stress decrease by increasing $m$. Figure 9b shows the tangential stress in this case, which has lower values relative to the previous state. In this case, in contrast to the $n = 0$ state, the tangential stresses have been converged to zero for all values of *m*, by approaching to end of the disc. Figure 9c represents the elongation along the disk, which increases with increasing *m*.

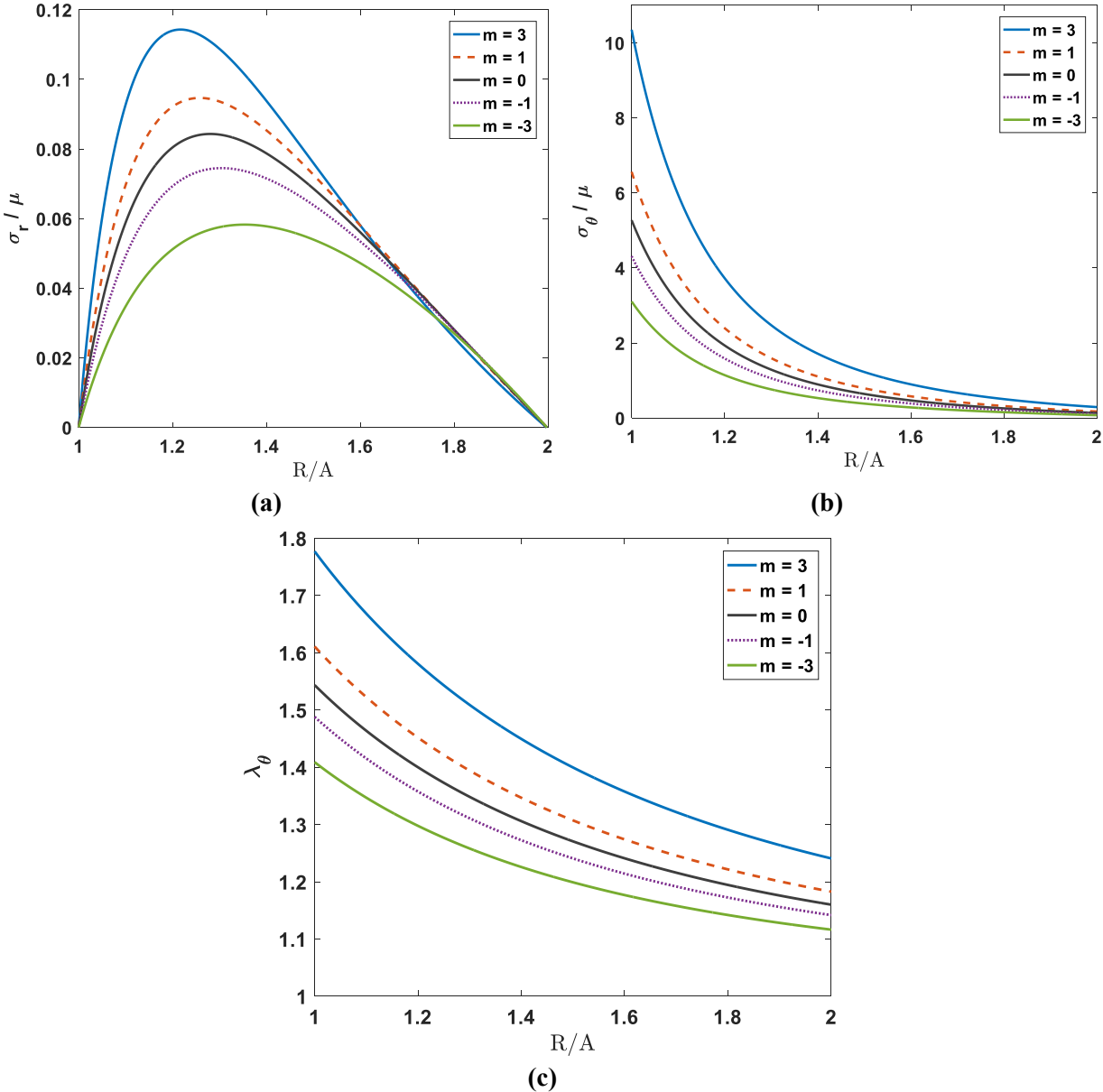

**Figure 9.** Results for *w* = 4.5 rad/s and *n* = -2 and different values of *m*, **(a)** radial stress distribution, **(b)** tangential stress distribution, **(c)** elongation.

In figure 10a, which indicates the distribution of radial stress, stress values are higher than the previous states. In this case as in the previous states, with increasing *m*, the maximum stress moves toward to the inner radius. But, compared to the previous state that the values of stresses had been decreasing with increasing *m* and in $\frac{R}{A} = 1.8$ until the end of disk, in this case, it would be obtained in $\frac{R}{A}=1.6$. Figure 10b shows the value of tangential stress that has increased compared to the previous state, and its value at the end of disk is non-zero and has almost a linear trend along the disk. Figure 10c represents the elongation along the disk, which increases with increasing *m*.

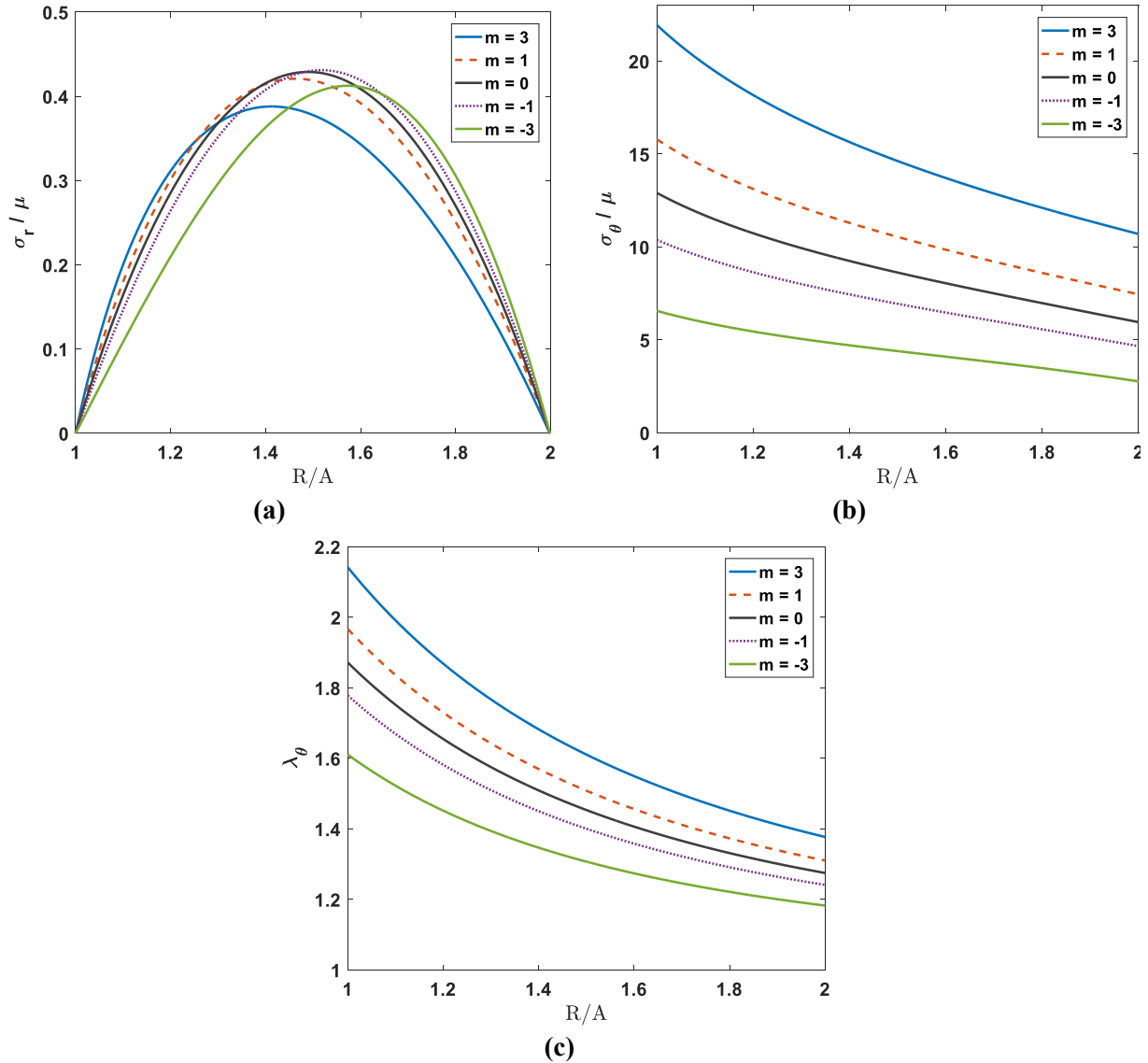

**Figure 10.** Results for $w = 4.5$ rad/s and $n = 2$ and different values of *m*, **(a)** radial stress distribution, **(b)** tangential stress distribution, **(c)** elongation.

### *3.3. Effects of angular velocity*

Figure 11a shows the radial stress in $m = 3$ and $n = -2$. The radial stress increases with increasing the rotational speed, but this increase does not affect the location of maximum radial stress. Figure 11b indicates the tangential stress distribution. As it is shown on this figure, the change in rotational speed has a high effect on the tangential stress. Figure 11c shows the elongation relative to the angular velocity, indicating an increase in elongation.

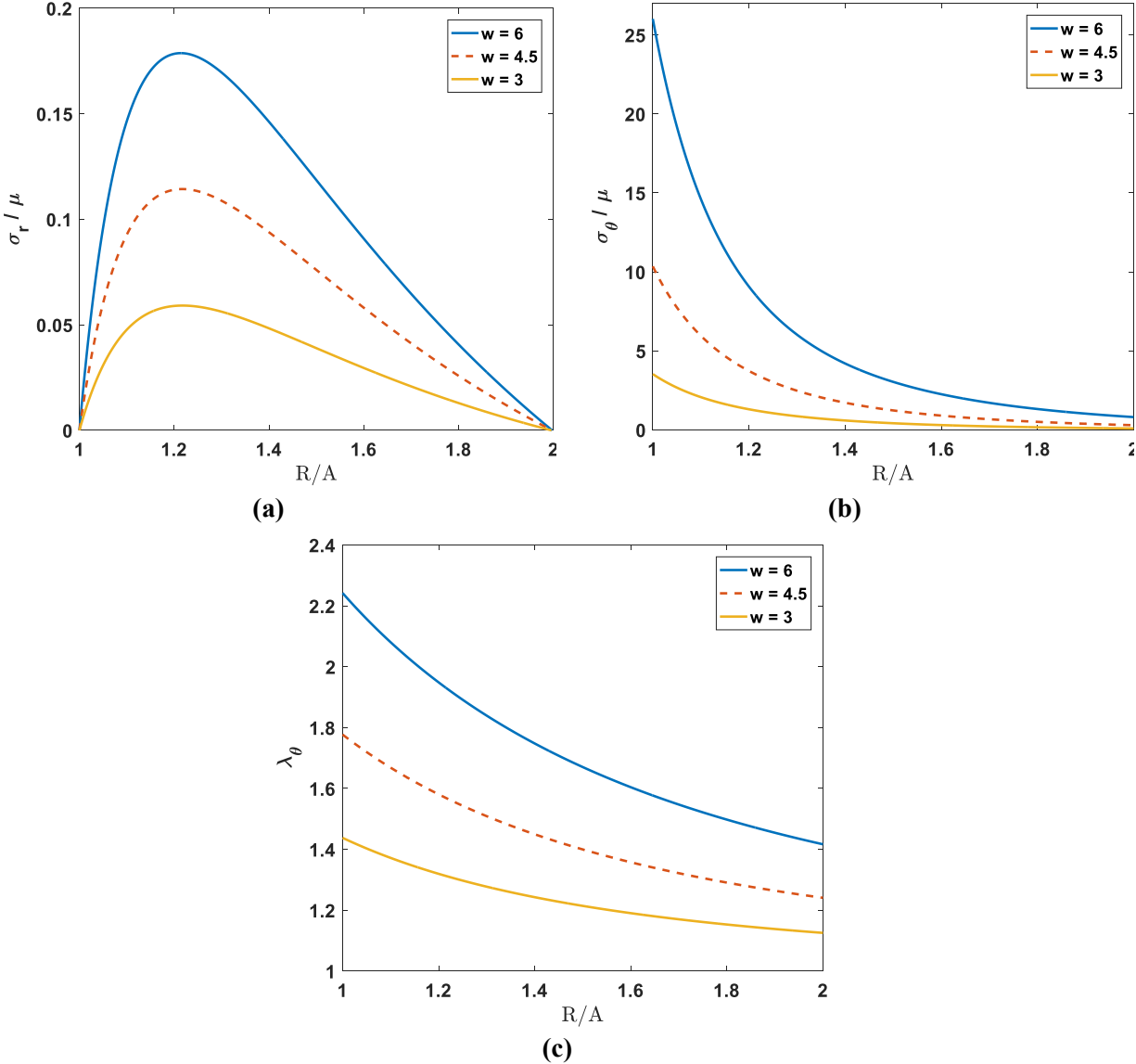

**Figure 11.** Results for *m* = 3 and *n* = -2 and different values of *w* (rad/s), **(a)** radial stress distribution, **(b)** tangential stress distribution, **(c)** elongation.

## 4. Conclusion

This research investigated the stress analysis of functionally graded hyperelastic variable thickness rotating annular thin disk. The thin disk in this study was made with variable properties in the radial direction. The variations of thickness and heterogeneity in disk were considered as a power relation with positive and negative coefficients. The performed analyzes have been considered for three different rotational speeds; and then the radial stress distribution, tangential stress distribution, and elongation were extracted. Results showed that increasing the angular velocity of the disk *w* leads to higher levels of radial stress, tangential stress, and elongation, with tangential stress experiencing a more significant rise. Moreover, in a constant cross section *m*, by increasing the value of heterogeneity coefficient *n*, the radial stresses increase and the maximum stress moves toward the outer radius. In this case, tangential stress increases with increasing *n*, and the distribution of tangential stress is such that it converges to small quantities for negative values of material properties. Additionally, the elongation increases with increasing *n*. For different values of *n*, there is a greater difference between the elongations in the inner radius of the disc relative to the outer radius. Another outcome showed by this work is that in a fixed heterogeneity *n*, by increasing the value of *m*, radial stress increases until a certain point, and then the radial stress decreases. Furthermore, by increasing the value of *m*, the maximum radial stress moves toward the inner radius. Increasing *m* also raises the tangential stress, but the trend of this increase can vary with respect to the value of *n*.

In this study, it has been shown that by using the variable cross section and heterogeneity of a hyperelastic material, different behaviors were observed in large deformations. Therefore, these FG materials can be employed in a wide range of applications. For instance, by selecting optimal values for different values of *n* and *m*, the location of maximum radial stress can be controlled in large deformations, which can be valuable in terms of structure design and engineering.

## Authors contribution statement

**Ehsan Jebellat:** Conceptualization, Methodology, Software, Formal analysis, Investigation, Writing – original draft, Supervision. **Iman Jebellat:** Methodology, Software, Formal analysis, Investigation, Visualization, Writing – review & editing.

## Declaration of Competing Interest

The authors declare that they have no known competing financial interests or personal relationships that could have appeared to influence the work reported in this paper.

## Data availability

Data will be made available on reasonable request.